\documentclass[twocolumn,amsmath,showpacs,nofootinbib]{revtex4-1}
\usepackage{graphicx}
\usepackage{dcolumn}
\usepackage{bm}
\usepackage{color} 
\usepackage{slashed}
\begin{document}
\newcommand{\hs}{\hspace*{0.5cm}}
\newcommand{\vs}{\vspace*{0.5cm}}
\newcommand{\be}{\begin{equation}}
\newcommand{\ee}{\end{equation}}
\newcommand{\bea}{\begin{eqnarray}}
\newcommand{\eea}{\end{eqnarray}}
\newcommand{\ben}{\begin{enumerate}}
\newcommand{\een}{\end{enumerate}}
\newcommand{\bde}{\begin{widetext}}
\newcommand{\ede}{\end{widetext}}
\newcommand{\nn}{\nonumber}
\newcommand{\crn}{\nonumber \\}
\newcommand{\Tr}{\mathrm{Tr}}
\newcommand{\non}{\nonumber}
\newcommand{\noi}{\noindent}
\newcommand{\al}{\alpha}
\newcommand{\la}{\lambda}
\newcommand{\bet}{\beta}
\newcommand{\ga}{\gamma}
\newcommand{\va}{\varphi}
\newcommand{\om}{\omega}
\newcommand{\pa}{\partial}
\newcommand{\+}{\dagger}
\newcommand{\fr}{\frac}
\newcommand{\bc}{\begin{center}}
\newcommand{\ec}{\end{center}}
\newcommand{\Ga}{\Gamma}
\newcommand{\de}{\delta}
\newcommand{\De}{\Delta}
\newcommand{\ep}{\epsilon}
\newcommand{\varep}{\varepsilon}
\newcommand{\ka}{\kappa}
\newcommand{\La}{\Lambda}
\newcommand{\si}{\sigma}
\newcommand{\Si}{\Sigma}
\newcommand{\ta}{\tau}
\newcommand{\up}{\upsilon}
\newcommand{\Up}{\Upsilon}
\newcommand{\ze}{\zeta}
\newcommand{\ps}{\psi}
\newcommand{\Ps}{\Psi}
\newcommand{\ph}{\phi}
\newcommand{\vph}{\varphi}
\newcommand{\Ph}{\Phi}
\newcommand{\Om}{\Omega}
\newcommand{\AdrHEPC}{Phenikaa Institute for Advanced Study and Faculty of Basic Science, Phenikaa University, Yen Nghia, Ha Dong, Hanoi 100000, Vietnam}

\title{Novel imprint of a vector doublet} 

\author{Phung Van Dong} 
\email{Corresponding author.\\ dong.phungvan@phenikaa-uni.edu.vn}
\affiliation{\AdrHEPC} 
\author{Duong Van Loi}
\affiliation{\AdrHEPC} 
\author{Le Duc Thien}
\affiliation{Department of Natural Science, Hoang Van Thu Specialized High School, Thinh Lang, Hoa Binh 350000, Vietnam}
\author{Pham Ngoc Thu}
\affiliation{Faculty of Mathematics-Physics-Informatics, Tay Bac University, Quyet Tam, Son La 360000, Vietnam}  

\date{\today}

\begin{abstract}
It is shown that the presence of a vector doublet is suitable to address neutrino mass, dark matter, and the recent muon anomalous magnetic moment.          
\end{abstract} 

\maketitle

\section{Motivation}

Neutrino mass and dark matter are the two leading questions in science, which require the new physics beyond the conventional standard model \cite{Zyla:2020zbs}. Additionally, the muon anomalous magnetic moment recently measured also reveals a significant deviation at 4.2$\sigma$ from the standard model prediction \cite{Abi:2021gix}.   

Within the attempts to solve simultaneously the first two issues, the scotogenic setup \cite{Ma:2006km} is the most compelling. Indeed, it introduces an inert scalar doublet and three sterile neutrinos, which couple to the usual lepton doublets, producing the relevant neutrino masses through the one-loop contribution of these new fields, while the lightest of which, either a neutral inert scalar or a sterile neutrino, is viably to be a dark matter candidate. However, an exact $Z_2$ parity, that necessarily makes all the new fields odd, plays the crucial role for the model properly working as well as stabilizing the dark matter. The origin of the $Z_2$ was left as an open question. Further, it is found that the minimal scotogenic contributes insignificantly (and negative) to the muon $g-2$ \cite{Chen:2019nud}. 

We suggest that if a new vector doublet is proposed instead of the inert scalar doublet, it cannot develop a vacuum expectation value due to the Lorentz invariance, in opposition to the inert scalar case that potentially breaks the $Z_2$, inducing an unwanted tree-level neutrino mass. Additionally, the standard model symmetry demands that the vector doublet couples to normal matter {\it only} through new fermions; for such reason, the sterile neutrinos are presented.\footnote{Indeed, the alternative choice includes new fermion triplets or (irrelevant) doublets, but not interested, since they are not universally supported by the gauge completion models.} Furthermore, the Lagrangian of the vector doublet realizes a matter parity, instead of the $Z_2$, which commonly arises as a residual gauge symmetry in the theories that let the vector doublet be a part of the extended gauge field. 

With the effective theory at hand, we give a novel recognition of a scotogenic mechanism with the vector doublet, leading to the neutrino mass, dark matter, and the muon $g-2$ manifestly. The new physics predicted is appropriate to the collider bounds. A brief discussion on the UV-completion theory is delivered at the end.

\section{Proposal} 

We introduce a vector doublet to the standard model, \be 
V_\mu=\left(\begin{array}{c}
V^0_\mu \\ V^-_\mu 
\end{array}\right)\sim (1,2,-1/2),\ee which has the $SU(3)_C\otimes SU(2)_L\otimes U(1)_Y$ quantum numbers like the ordinary lepton doublets, $\psi_{aL}=(\nu_{aL}\ e_{aL})^T\sim(1,2,-1/2)$, where $a=1,2,3$ is a family index. Notice that $\tilde{V}_\mu \equiv i\sigma_2 V^*_\mu \sim (1,2,1/2)$ transforms as the usual Higgs doublet, $\phi=(\phi^+\ 
\phi^0)^T \sim (1,2,1/2)$, and the introduction of either $V$ or $\tilde{V}$ from outset is physically equivalent. Such a vector doublet is commonly hinted from numerous electroweak gauge extensions, such as the 3-3-1 \cite{Singer:1980sw,Montero:1992jk,Foot:1994ym}, trinification \cite{Babu:1985gi}, $SU(6)$ and $E_6$, as well as gauge-Higgs unification \cite{Fairlie:1979at,Hosotani:1983xw}, that all contain the higher weak isospin, $SU(3)_L$, as a sub- or residual group~\cite{Chizhov:2009fc}.

The Lagrangian of the vector doublet according to the standard model symmetry takes the form, \bea \mathcal{L} &\supset& -\fr 1 2 F^\dagger_{\mu\nu} F^{\mu \nu} +m^2_V V^\dagger_\mu V^\mu -\fr{1}{\xi}(D_\mu V^\mu)^\dagger (D_\nu V^\nu)\crn
&&+ \al_1 (V^\dagger_\mu V^{\mu}) (V^\dagger_\nu V^{\nu}) + \al_2 (V^\dagger_\mu V^{\nu}) (V^\dagger_\nu V^{\mu})\crn
&& +\al_3 (V^\dagger_\mu V^{\nu}) (V^{\dagger \mu} V_{\nu}) + i \beta_1 V_\mu^\dagger A^{\mu\nu} V_\nu + i \beta_2 V_\mu^\dagger B^{\mu\nu} V_\nu
\crn
&& + \la_1 (\phi^\dagger\phi) (V^\dagger_\mu V^{\mu}) + \la_2(\phi^\dagger V_{\mu})(V^{\dagger \mu}\phi)\crn
&& +\fr 1 2 [ \la_3(\phi V_{\mu})(\phi V^{\mu})+H.c. ],\label{lb1}
\eea where $F_{\mu\nu}=D_\mu V_\nu - D_\nu V_\mu$, $D_\mu=\pa_\mu + ig A_{\mu} + i g' B_\mu$, and $A_{\mu}$ ($A_{\mu\nu}$) and $B_\mu$ ($B_{\mu\nu})$ are the gauge field (field strength) of $SU(2)_L$ and $U(1)_Y$, respectively.\footnote{Here $\phi V = \phi^+ V^--\phi^0 V^0\sim 1$ under $SU(2)_L$. The couplings of other forms, e.g. $(\phi V)(\phi V)\sim 3\times 3$ and $(\phi V)^\dagger (\phi V)\sim 1\times 1$ or $3^*\times 3$, are reducible to the given interactions, thus suppressed.} Notice that the nontrivial $1/\xi$ term is simply an allowed gauge-invariant coupling, not a gauge fixing, since $V_\mu$ is not a gauge field. Surely from (\ref{lb1}), the vector fields obey the Klein-Gordon equation as well as the Lorentz condition manifestly. One would replace $A\rightarrow T_j A_j$ and $B\rightarrow Y B$ for practical computation, in which $T_j$ ($j=1,2,3$) and $Y$ are the weak isospin and hypercharge, respectively.  

This theory preserves an accidental $Z_2$ symmetry, which transforms $V_\mu\rightarrow -V_{\mu}$, while leaving the standard model fields unchanged, in agreement with \cite{Saez:2018off}. The $Z_2$ symmetry is identical to the matter parity \be P=(-1)^{3(B-L)+2s}\ee to be a residual gauge symmetry of the 3-3-1-1 gauge group \cite{Dong:2013wca} or flipped trinification \cite{Dong:2017zxo} breaking, there the vector doublet has $B-L=-1$. Unlike all other couplings, the $\la_3$ interaction violates $B-L$, but conserves $P$, to be induced by breaking $B-L=2$. That said, it should be radically smaller than the $B-L$ conserving couplings, such as \be \la_3\ll \la_{1,2},\al_{1,2,3}, \beta_{1,2},1/\xi.\ee On the other hand, the scalar-vector mixing terms of type, $i\kappa \phi D_\mu V^\mu+H.c.$, disappear due to $Z_2$ conservation, or they are automatically eliminated if $V_\mu$ originates as a part of gauge field of a higher gauge symmetry.  

The vector doublet does not couple to the standard model fermions due to $\bar{f}_L \ga^\mu f_R X_\mu=0$ for $X=V$ or $\tilde{V}$ as suitable. However, we introduce three fermion singlets, \be N_{aL} \sim (1,1,0),\ee to be the variants of the standard model neutrinos. The fermion singlets couple to the standard model lepton doublets through the vector doublet,
\be \mathcal{L}\supset h_{ab} \bar{\psi}_{aL}\ga^\mu N_{bL} V_\mu +H.c.\label{lb2}\ee  The new fermions transform under $Z_2$ as $N_{aL}\rightarrow -N_{aL}$, or they have $B-L=0$ under the matter parity, and they possess mass terms, \be \mathcal{L}\supset -\fr 1 2 M_{ab} N_{aL} N_{bL}+H.c.\label{lb3}\ee Here we assume $M=\mathrm{diag}(M_1,M_2,M_3)$ to be flavor diagonal, without loss of generality. We label the physical states and masses, say $N_k, M_k$, in the mass basis.    

After electroweak symmetry breaking, the vector doublet components $V^0_{\mu}\equiv (V_{1\mu}+i V_{2\mu})/\sqrt{2}$ and $V^-_\mu$ are separated in mass, such as \bea m^2_{V^-} &=& m^2_V+(\la_1+\la_2)\fr{v^2}{2},\\
m^2_{V_1} &=& m^2_V+(\la_1+\la_3)\fr{v^2}{2},\\
m^2_{V_2} &=& m^2_V+(\la_1-\la_3)\fr{v^2}{2}.\eea Here $\la_3$ has been assumed to be real; otherwise, its phase can be removed by redefining appropriate $\phi,V$ fields. Moreover, the Feynman propagators of the vector fields are similar to massive gauge fields in a $R_\xi$ gauge due to the contribution of the $1/\xi$ coupling. On the other hand, the addition of the vector fields violates the unitarity condition of the $S$-matrix and the theory only works well below some cut-off scale, $\La \sim m_V$, making constraint on the new physics couplings, naively \be \fr{g^2_{\mathrm{NP}}}{4\pi} \lesssim \fr{\La^2}{s}\sim 1.\ee We take $\La\sim \sqrt{s}\sim 1$ TeV to be the current energy of colliders, where the standard model is still good.    

The addition of $V_\mu$ and $N_{aL}$ presents three important results, arranged in order.  

\section{Scotogenic scheme} 

The scotogenic model \cite{Ma:2006km} is an idea of radiative-seesaw neutrino-mass generation, through one-loop diagram mediated by an inert Higgs doublet and three sterile neutrinos. It is recognized in this work in which the doublet vector plays the role of the inert Higgs doublet instead, with the similar interactions such as $\la_3$ in (\ref{lb1}) and $h_{ab}$ in (\ref{lb2}), including the Majorana nature of $N_{aL}$ in (\ref{lb3}).\footnote{Notice that Ref. \cite{CarcamoHernandez:2018vdj} introduced, by contrast, highly nonrenormalizable (dimension-12) operators connecting $V_\mu$ to leptons in a theory under discrete symmetries, without possessing the $1/\xi$ coupling or a gauge completion; as a result, the neutrino mass diagram was divergent, defined by a cutoff.}

\begin{figure}[h]
\includegraphics[scale=0.9]{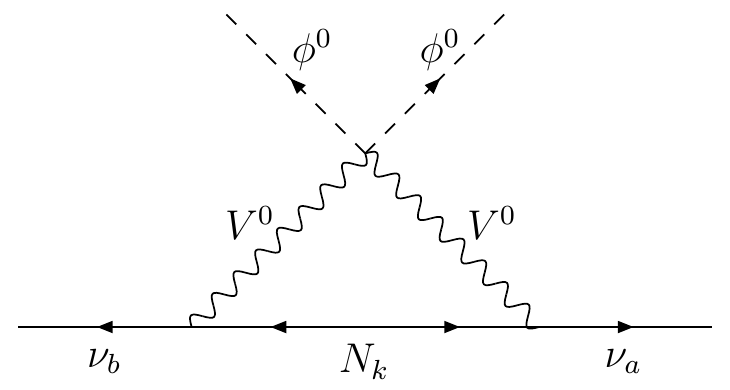}
\caption[]{\label{fig1} One-loop contribution to neutrino mass.}
\end{figure}
The neutrino mass generation diagram is depicted in Fig. \ref{fig1}. Hence, the neutrino mass matrix, that is induced in the form of $\mathcal{L}\supset -\fr 1 2 \bar{\nu}_{a L} (M_\nu)_{ab}\nu^c_{bL}$, reads 
\bde\bea -i (M_\nu)_{ab}P_R&=&\int\fr{d^4 p }{(2\pi)^4}\left(i\fr{h_{ak}}{\sqrt{2}}\ga^\mu P_L\right) \fr{i}{\slash\!\!\! p-M_k}
\left(-i\fr{h_{bk}}{\sqrt{2}}\ga^\nu P_R\right)\fr{-i}{p^2-m^2_{V_1}}\left(g_{\mu\nu}-\fr{(1-\xi)p_\mu p_\nu}{p^2-\xi m^2_{V_1}}\right)\crn
&&+\int\fr{d^4 p }{(2\pi)^4}\left(-\fr{h_{ak}}{\sqrt{2}}\ga^\mu P_L\right) \fr{i}{\slash\!\!\! p-M_k}
\left(\fr{h_{bk}}{\sqrt{2}}\ga^\nu P_R\right)\fr{-i}{p^2-m^2_{V_2}}\left(g_{\mu\nu}-\fr{(1-\xi)p_\mu p_\nu}{p^2-\xi m^2_{V_2}}\right).\eea \ede Notice that the real ($V_1$) and imaginary ($V_2$) parts of $V^0$ give opposite contributions, since the product of the two neutrino vertices to each of $V_{1,2}$ changes sign, and the difference between the contributions is only due to the $V_{1,2}$ mass splitting. It is straightforward to obtain,
\bea (M_\nu)_{ab} &=& \fr{h_{ak}h_{bk}M_k}{32\pi^2}\left[f(M^2_k,m^2_{V_1})-f(M^2_k,m^2_{V_2})\right],\eea where $f(x,y)$ equals to \bea \fr{y}{x-y}\left[\fr{(1+\xi)(3-\xi)x-\xi(3+\xi)y}{x-\xi y}\ln\fr{y}{x} -\fr{x-y}{x-\xi y}\xi^2\ln \xi \right].\nn\eea  

Assume \be m^2_{V_1}-m^2_{V_2}=\la_3 v^2\ll m^2_0\equiv m^2_V+ \la_1 v^2/2,\ee where $m^2_0 = (m^2_{V_1}+ m^2_{V_2})/2$. Additionally, assume \be \xi\gg \mathrm{Max}(M_{1,2,3}/m_0)^2.\ee We find 
\be (M_\nu)_{ab}\simeq -\fr{\la_3\xi}{32\pi^2} \fr{h_{ak}h_{bk}M_kv^2}{m^2_0}. \ee  The scale $m^2_0/M_k$ as in inverse seesaw is significantly reduced by the loop factor ($1/16\pi^2$) and the small coupling $(\la_3 \xi)$. Indeed, suppose Max$(M_{1,2,3})\sim 10$ GeV and $\la_3 \xi\sim 10^{-7}$, slightly smaller than the electron Yukawa coupling.\footnote{$\la_3$ is associated with the $B-L$ breaking scale that determines the matter parity \cite{Dong:2013wca,Dong:2017zxo}, so it is proportional to $\la_3\sim m_0/\La_{B-L}\lesssim 10^{-8}$, given that $\La_{B-L}\gtrsim 10^{11}$ GeV.} The neutrino masses are recovered, i.e. $M_\nu\sim 0.1$ eV, requiring only $h_{ak}h_{bk}\sim 1$ and $v/m_0\sim 1/3$, appropriate to the following dark matter density and muon $g-2$. Further, the charged leptons considered are flavor diagonal, while $\nu_a$ are related to the mass eigenstates $\nu_i$ by the PMNS matrix, $\nu_a=U_{ai} \nu_i$.      

\section{Dark matter}

\begin{figure}[h]
\includegraphics[scale=0.9]{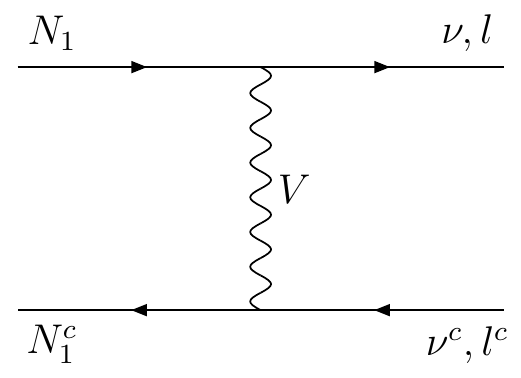}
\caption[]{\label{fig2} Dark matter annihilation to normal matter.}
\end{figure}

The dark matter candidate is the lightest field of either the fermion singlets $N_{1,2,3}$ or the neutral vectors $V_1, V_2$, stabilized by the accidental symmetry $Z_2$. The dark matter phenomenology of the vector candidate was discussed in \cite{Saez:2018off} in a model without $N_{1,2,3}$. It is noted that the fermion singlets $N_{1,2,3}$ may govern the vector dark matter observables through the interaction in (\ref{lb2}). This work does not consider such case, instead we interpret the lightest fermion singlet (assumed $N_1$, thus $M_1<M_{2,3},m_{V_{1,2}},m_{V^-}$) as dark matter. Its annihilation is given in the diagram in Fig. \ref{fig2}.       

The total annihilation cross-section, $N_1 N^c_1\rightarrow l' l^c (\nu_{l'} \nu^c_l)$, is evaluated as 
\be \langle \sigma v \rangle \simeq \fr{\sum_{l,l'}|h^*_{l1}h_{l'1}|^2}{8\pi M^{-2}_1}\left(1+\fr{8}{x_F}\right)\left(\fr{1}{m^4_{V^-}}+\fr{1}{m^4_{0}}\right),\ee where $M_1$ is the $N_1$ mass, $h_{l1}$ and $h_{l'1}$ for $l,l'=e,\mu,\tau$ are the $h$ couplings in ({\ref{lb2}}) that equally couple $N_1$ to a pair of charged leptons $(l l')$ or neutrino flavors $(\nu_l \nu_{l'})$, and $x_F=M_1/T_F\sim 20$--25 at the freeze-out temperature.   

Assuming $m_{V^-}\simeq m_0$, it leads to \bea \langle \sigma v\rangle &\simeq& 1\ \mathrm{pb}\left( \fr{\sum_l |h_{l1}|^2}{4\pi}\right)\left(\fr{\sum_{l'} |h_{l'1}|^2}{4\pi}\right)\crn
&&\times  \left(\fr{88 M_1}{m_0}\right)^2\left(\fr{800\ \mathrm{GeV}}{m_0}\right)^2,\eea where the benchmark value of $m_0$ is given in the following section. The dark matter gets the correct density if the couplings $\sum_l|h_{l1}|^2/4\pi$ are proportional to unity and the dark matter mass $M_1\sim m_0/88\sim 9$ GeV. 

\section{Muon $g-2$}

In the standard model, the anomalous magnetic moment of muon, $a_\mu=(g-2)_{\mu}/2$, receives the radiative contributions from the electromagnetic, hadronic, and electroweak parts, now established \cite{Aoyama:2020ynm}, \be a_{\mu}(\mathrm{SM})=116591810(43) \times 10^{-11}.\ee The recent measurement of $a_\mu$ presents an exciting hint for the new physics beyond the standard model \cite{Abi:2021gix}. The new result combined with the previous E821 measurement \cite{Bennett:2006fi} yields a deviation from the standard model prediction at 4.2$\sigma$: \be a_{\mu}(\mathrm{Exp})-a_{\mu}(\mathrm{SM})=(251\pm 59)\times 10^{-11}.\label{lb5}\ee  

If this deviation is further confirmed, it might rule out many new physics models, due to the fact that the deviation is bigger than the electroweak contributions of the standard model coming from $W,Z,H$ mediators, i.e. $a_\mu(\mathrm{EW})=153.6(1.0)\times 10^{-11}$. Indeed, the new physics constrained is potentially in tension with the electroweak precision test and the other collider bounds. 

\begin{figure}[h]
\includegraphics[scale=0.9]{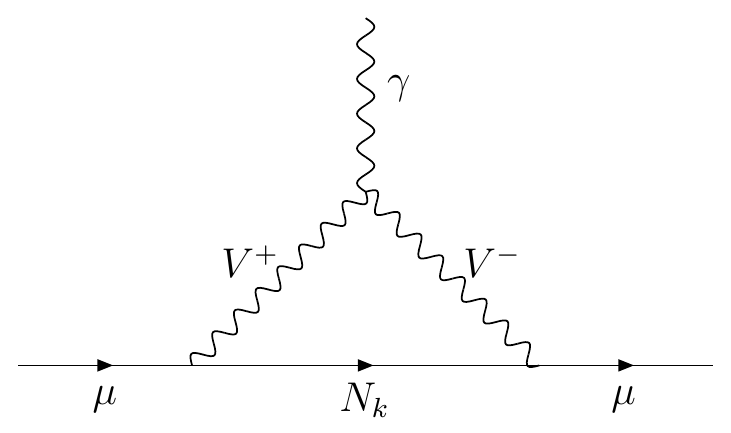}
\caption[]{\label{fig3} Doublet vector contribution to the muon $g-2$.}
\end{figure}
We suggest to solve this question by a diagram exchanged by the doublet vector depicted in Fig. \ref{fig3}. Assuming $m_\mu\ll m_{V^-},M_k$, one obtains 
\bea \Delta a_\mu &=& \fr{1}{8\pi^2}\fr{m^2_\mu}{m^2_{V^-}}\sum_k |h_{\mu k}|^2\crn
&\times& \int^1_0 dx x \fr{x(1+x)m^2_{V^-}+(1-x)(1-\fr{x}{2})M^2_k}{x m^2_{V^-}+(1-x)M^2_k}.\eea The integral is proportional to 1 for $m_{V^-}\gg M_k$. Hence,
\be \Delta a_\mu\simeq 2.5\times 10^{-9}\left(\fr{800\ \mathrm{GeV}}{m_{V^-}}\right)^2\left(\fr{\sum_k |h_{\mu k}|^2}{4\pi}\right),\ee comparable to the muon $g-2$ in (\ref{lb5}), given that $m_{V^-}\simeq 800$ GeV and $\sum_k |h_{\mu k}|^2/4\pi\simeq 1$ as the unitarity bound.

\section{Summary and remark}

Unlike the photon and other gauge fields, massive vector fields are always quantized properly with three physical degrees of freedom due to automatic Lorentz condition and a propagator analogous to that in a gauge theory with unitary gauge. The standard model can contain such a vector doublet with a coupling to the electroweak gauge bosons, $\fr{1}{\xi}|D^\mu V_\mu|^2$, the Higgs field, $\la_3(\phi V_\mu)^2$, and the three singlet variants of light neutrinos, $h\bar{\psi}_L \ga^\mu N_L V_\mu$, but not to ordinary fermions. Here $V_\mu$ and $N_L$ are dark, ensured by the $Z_2$ or $P$ matter parity. 

The scotogenic producing small neutrino mass is recognized with the exchange of  the dark neutral vector $V^0$ and the dark fermion $N$. Whereas, a similar scotogenic producing the observed muon $g-2$ is realized with the mediator of the dark charged vector $V^\pm$ and the dark fermion $N$, given that $m_V\sim 800$ GeV. The lightest fermion singlet $N_1$ properly contributes to the dark matter abundance if its mass is $M_1\sim 9$ GeV. 

The fermion dark matter does not interact with nucleons in the direct detection experiments, while it can scatter with electrons in the XENON1T experiment \cite{Aprile:2020tmw}, but gives a tiny recoil energy $2m_e v^2\sim 1 $ eV as well as small signal strength as suppressed by $m^2_e/m^4_0$. The predicted vector doublet mass $m_V\sim 800$ GeV would satisfy all the high energy collider bounds, given that the $B-L$ conserving couplings in (\ref{lb1}) are perturbative, as shown in \cite{Saez:2018off}. So we do not refer to such constraints here. 

On the other hand, a mono-photon event may be created at the LEPII recoiled against the missing energy carried by a pair of dark matter $N_1 N_1$, governed by the effective interaction, \be \fr{|h_{e1}|^2}{m^2_V}(\bar{e}_L \ga^\mu N_{1L})(\bar{N}_{1L} \ga_\mu e_L).\ee Using the Fierz identities, we can translate this interaction to the vector and axial vector operators studied in Ref. \cite{Fox:2011fx}, implying a bound on $m_V>(450$--$500)\sqrt{\pi}=797$--886 GeV, for $|h_{e1}|^2/4\pi\sim 1$, which is in good agreement with the relic density and muon $g-2$ bound in this model. 

A crucial difference between our approach and potential $U(1)$ gauge extensions is that the vector doublet by its non-Hermitian nature behaves as a charged current linking distinct flavors, even in a gauge completion, in similarity to the $W$ boson. Whereas, the Abelian gauge fields often conserve flavors, hence there are a few parameters in such theories, such as the gauge boson mass, kinetic mixing, and coupling constant. As a result, a portion of them may encounter tensions in predicting all the mentioned experiments, although they are very predictive \cite{Bauer:2018onh}.

The UV-completion is related to the search for a gauge symmetry that contains the doublet $V_\mu\sim 2$ in its adjoint gauge representation. The smallest of which is the 3-3-1-1 gauge symmetry \cite{Dong:2013wca}, because the $SU(3)_L$ adjoint is decomposed as $8=3\oplus 2\oplus 2^*\oplus 1$ under $SU(2)_L$, which includes $V_\mu$ and $V^*_\mu$, as mentioned. In this case, $SU(3)_L$ is broken by a new Higgs boson unified with a Goldstone boson doublet of $V_\mu$ in a multiplet. Additionally, the fermion singlets are unified with the usual lepton doublets in $SU(3)_L$ triplets, such as $(\nu_{aL}\ e_{aL}\ N_{aL})^T$. $N_{aL}$ may have respective right-handed partners, $N_{aR}$, and all the fields, $N$'s and $V$'s, are odd under the matter parity as a residual 3-3-1-1 symmetry \cite{Dong:2015yra}. All the couplings in Eq. (\ref{lb1}) are restricted by the gauge symmetry and the model is now predictive. Particularly, the $\la_3$ coupling is induced by the gauge interaction of a scalar multiplet, e.g. an octet, that contains a Higgs doublet. Although $h_{ab}$ is restricted to the gauge coupling, the flavor change happens due to a mixing of $N_{aL}$, which shifts $h_{ak}\rightarrow (g/\sqrt{2})V_{ak}$ relating $N_{aL}$ to mass eigenstates $N_{k}$. Here note that the $N_{aL}$ mass can be induced by a seesaw, with the Dirac mass connecting $N_L$ to $N_R$ given by a $SU(3)_L$ breaking, while $N_R$ has a large Majorana mass by itself. The $1/\xi$ coupling becomes a gauge fixing canceled out by the contribution of the Goldstone boson modes of $V_\mu$ in the $R_\xi$ gauge for relevant physical processes. It is noteworthy that in this gauge completion, the inert scalar doublet in the usual scotogenic sense becomes the Goldstone modes of $V_\mu$, hence the vector doublet governs the neutrino mass and other observables instead. All this is out of the scope of the work, to be published elsewhere.                            

\section*{Acknowledgments}
This research is funded by Vietnam National Foundation for Science and Technology Development (NAFOSTED) under grant number 103.01-2019.353.

\bibliographystyle{JHEP}
\bibliography{combine}

\end{document}